\documentstyle[preprint,oldlfont,eqsecnum,aps]{revtex}
\begin{document}
\preprint{U. C. Irvine Technical Report 98-14}
\title{The Full Range of Predictions for
{\boldmath $B$} Physics \\ 
From Iso-singlet Down Quark Mixing}
\author{Dennis Silverman}
\address{Department of Physics and Astronomy, \\
University of California, Irvine \\
Irvine, CA 92697-4575 }
\date{\today}
\maketitle

\begin{abstract}
We extend the range of predictions of the isosinglet (or vector)
down quark model to the fully allowed physical ranges, and also update
this with the effect of new physics constraints.  We constrain the
present allowed ranges of $\sin(2\beta)$ and $\sin(2\alpha)$,
$\gamma$, $x_s$, and $A_{B_s}$.  In models allowing mixing to a new
isosinglet down quark (as in E$_6$) flavor changing neutral currents
are induced that allow a $Z^0$ mediated contribution to $B-\bar B$
mixing and which bring in new phases.  In $(\rho,\eta)$,
$(x_s,\sin{(\gamma)})$, and $(x_s, A_{B_s})$ plots for the extra
isosinglet down quark model which are herein extended to the full physical
range, we find new allowed regions that will require experiments on
$\sin{(\gamma)}$ and/or $x_s$ to verify or to rule out an extra down
quark contribution.
\end{abstract}
\pacs{11.30.Er,12.15.Hh,12.15.Mm,12.60-i,14.40.Nd}

\section{Introduction}
The ``new physics'' class of models we use are those with extra
iso-singlet down quarks, where we take only one new down quark as
mixing significantly.  An example is E$_6$, where there are two down
quarks for each generation with only one up quark, and of which we
assume only one new iso-singlet down quark mixes strongly.  This model
has shown large possible effects in $B-\bar B$ mixing phases.  The
approaching $B$ factory experiments will also sets limits on the
phases of the mixing angles to the new iso-singlet down quark in this model.  
In previous analyses\cite{chosilckm,sil95}, 
we focused on ranges of variables in which the
standard model (SM) results occurred, in the sense of looking for small
deviations in setting limits.  As emphasized by 
Wolfenstein\cite{wolfenstein}, we
now explore the full range of output in variables $\eta$,
$\sin{(\gamma)}$, and the $B_s$ asymmetry to indicate the full possible
range of outcomes for these experiments due to new physics models.

A significant number of improved constraints have appeared in the last two
years, and most importantly, some of the $R_b$ experiments now give
results in agreement with the standard model.  Since the mixing to a
new down quark can only decrease the diagonal neutral current, these
results now give useful limits on the parameters.  The other improved
experiments are $K^+ \to \pi^+ \nu \bar{\nu}$, the new D0 limit on $B
\to \mu \mu X$, improved $V_{ub}$ limits, and the LEP lower bounds on
$\Delta m_s$ or $x_s$.  We also now have an exact method of combining
the one event Poisson result on $K^+ \to \pi^+ \nu \bar{\nu}$ with the
Gaussian probability experiments which results in a chi-squared
distribution\cite{silstats}.

We also project to a range of results from the $B$ factory experiments.
For different $\sin({2\alpha})$ cases we find extended multiple regions
in $(\rho, \eta)$
that will require experiments on $\sin{(\gamma)}$ or $x_s$ to decide
between, and experiments on both could be required to effectively
bound out or to verify the model.  We also find a sizeable range for the 
$B_s - \bar{B}_s$ mixing asymmetry in the extra down quark model, 
while in the SM this asymmetry is very small.  
In setting limits we use the method of a joint $\chi^2$ fit to all
constraining experiments.

\section{Iso-singlet Down Quark Mixing Model}

Groups such as $E_6$ with extra SU$(2)_L$ singlet down quarks give
rise to flavor changing neutral currents (FCNC) through the mixing of
four or more down quarks \cite{sil95,shin,nirsilnp,nirsilpr,sil92}.
We use the $4 \times 4$ down quark mixing matrix $V$ which
diagonalizes the initial down quarks ($d_{iL}^0$) to the mass
eigenstates ($d_{jL}$) by $d_{iL}^0 = V_{ij} d_{jL}$.  The flavor
changing neutral currents we have are \cite{nirsilpr,sil92} $-U_{ds} =
V^*_{4d} V_{4s}$ , $-U_{sb} = V^*_{4s} V_{4b}$, and $-U_{bd} =
V^*_{4b} V_{4d}$.  These FCNC with tree level $Z^0$ mediated exchange
may contribute part of $B_d^0 - \bar{B_d^0}$ mixing and of $B_s^0 -
\bar{B_s^0}$ mixing, and the constraints leave a range of
values for the fourth quark's mixing parameters.  $B_d^0 -
\bar{B_d^0}$ mixing may occur by the $b - \bar{d}$ quarks in a
$\bar{B_d}$ annihilating to a virtual $Z$ through a FCNC with
amplitude $U_{db}\ $, and the virtual $Z$ then creating $\bar{b} - d$
quarks through another FCNC, again with amplitude $U_{db}$, which then
becomes a $B_d$ meson.  If these are a large contributor to the $B_d
-\bar{B_d}$ mixing, they introduce three new mixing angles and two new
phases over the standard model (SM) into the $CP$ violating $B$ decay 
asymmetries.
The size of the
contribution of the FCNC amplitude $U_{db}$ as one side of the
unitarity quadrangle is less than 0.15 of the unit base $|V_{cd}
V_{cb}|$ at the 1-$\sigma$ level, but we have found
\cite{sil95,shin,nirsilpr,sil92} that it can contribute, at present,
as large an amount to $B_d -\bar{B}_d$ mixing as does the standard
model.  The new phases can appear in this mixing and give total phases
different from that of the standard model in $CP$ violating $B$ decay
asymmetries\cite{nirsilpr,sil92,chosilfcnc,branco,lavoura}.

For $B_d - \bar{B}_d$ mixing with the four down quark induced $b-d$
coupling, $U_{db}$, we have \cite{chosilfcnc}
\begin{equation} 
x_d = (2 G_F/3 \sqrt{2}) B_B f_B^2 m_B \eta_B \tau_B \left|
U_{std-db}^2 + U_{db}^2 \right| 
\end{equation}
where with $y_t = m_t^2/m_W^2$
\begin{equation}
U^2_{std-db} \equiv  (\alpha/(4 \pi \sin^2{\theta_W})) y_t
f_2(y_t) (V_{td}^* V_{tb})^2,
\end{equation} 
and $x_d = \Delta
m_{B_d}/\Gamma_{B_d} = \tau_{B_d} \Delta
m_{B_d}$.

The $CP$ violating decay asymmetries depend on the combined phases of
the $B^0_d-\bar{B}^0_d\;$ mixing and the $b$ quark decay amplitudes
into final states of definite $CP$.  Since we have found that $Z$
mediated FCNC processes may contribute significantly to
$B^0_d-\bar{B}^0_d\;$ mixing, the phases of $U_{db}\ $ would be
important.  Calling the singlet down quark $D$, to leading order the
mixing matrix elements to $D$ are $V_{tD} \approx s_{34}$, $V_{cD}
\approx s_{24} e^{-i\delta_{24} }$, and $V_{uD} \approx s_{14} e^{-i
\delta_{14} }$.  The complete $4 \times 4$ mixing matrix was given
previously\cite{chosilfcnc,wang}.
The FCNC amplitude $U_{db}$ to leading order in the
new angles is
\begin{equation}
U_{db} = (-s_{34} - s_{24} s_{23} e^{i\delta_{24}})
(s_{34} V^*_{td} + s_{14} e^{-i \delta_{14} } -s_{24}
e^{-i \delta_{24} } s_{12}).
\end{equation}
where $V_{td} \approx (s_{12} s_{23} - s_{13} e^{i\delta_{13}})$, and
$V_{ub} = s_{13} e^{-i\delta_{13}}$.

\section{Joint Chi-squared Analysis for CKM and FCNC Experiments}

FCNC experiments put limits on the new mixing angles and constrain the
possibility of new physics contributing to $B_d^0 - \bar{B_d^0}$
and $B_s^0 - \bar{B_s^0}$ mixing.  Here we jointly analyze all
constraints on the $4 \times 4$ mixing matrix obtained by assuming
only one of the SU$(2)_L$ singlet down quarks mixes
appreciably\cite{nirsilpr}.  We use the nine experiments for the $3
\times 3$ CKM sub-matrix elements \cite{chosilckm}, which include:
those on the five matrix elements $V_{ud}, V_{cd}, V_{us}, V_{ub},
V_{cb}$ of the $u$ and $c$ quark rows; $|\epsilon|$ and $K_L \to \mu \mu$  
in the neutral $K$ system\cite{detail}; $B_d-\bar{B}_d$ mixing ($x_d$);
and the new limits on $\Delta m_s$, or $x_s$.  
For studying FCNC, we have four experiments which include the bound on 
$B \to \mu \mu X$ (which constrains $b \to d$ and $b \to s$) for which
we have 
the UA1 and the new D0\cite{D0} results, the new first event in $K^+ \to
\pi^+ \nu \bar{\nu} $\cite{silstats,lavoura,kpiexpt,newkpi,onekpi} 
and new results on $R_b$ in $Z^0 \to b \bar{b}$ \cite{lavoura,rbexpt} (which
directly constrains the $V_{4b}$ mixing element).  FCNC experiments
will bound the three amplitudes $U_{ds}$, $U_{sb}$, and $U_{bd}$ which
contain three new mixing angles and three phases.  We use the mass of
the top quark as $m_t = 174$ GeV.  We use a method for combining 
the Bayesian Poisson distribution for the average for the 
one observed event in $K^+ \to \pi^+ \nu \bar{\nu}$\cite{onekpi}
with the chi-squared distribution from the other experiments.  We take
$<n> = 2.7 \times 10^8 ~ |U_{ds}|^2$, ignoring the SM contribution since the
observed event is at a rate four times the SM result.

In maximum likelihood correlation plots, we use for axes two output
quantities which are dependent on the mixing matrix angles and phases, 
such as $(\rho,\eta)$, and
for each possible bin with given values for these, we
search through the nine dimensional angular data set of the $4 \times
4$ down quark mixing angles and phases, finding all sets which give results in
the bin, and then put into that bin the minimum $\chi^2$ among
them.  To present the results, we then draw contours at several
$\chi^2$ in this two dimensional plot corresponding to given confidence levels.

\section{Constraints on the Standard Model CKM Matrix at Present}

We first analyze the standard model using the present constraints on
the eight CKM related experiments.  We use the results for
$|V_{ub}/V_{cb}| = 0.08 \pm 0.016$ or a 20\% error\cite{ali}.

In Fig.~\ref{rho-etaSM} is shown the $(\rho,\eta)$ plot for the
standard model with contours at $\chi^2$ which correspond to
confidence levels (CL) that are the same as the CL for 1-$\sigma$,
2-$\sigma$, and 3-$\sigma$ limits.  Fig.~\ref{rho-etaSM} shows large
regions for the present CKM constraints.  We see the effects of the
$x_s = 1.35 x_d |V_{ts}/V_{td}|^2$ lower bound in the SM limiting the 
length of $V_{td} \propto \sqrt{(1-\rho)^2+\eta^2}$ and starting to 
cut off $\rho$ for $\rho < 0$. 

In Fig.~\ref{sa-sbSM} is shown the $(\sin{(2\alpha)},\sin{(2\beta)})$
plot for the standard model, for the same cases as in Fig. 1.  In
comparison to previous analyses\cite{sil95} the region near
$\sin{(2\alpha)}=1$ is no longer within the 1-$\sigma$ contour.

In Fig.~\ref{xs-sgSM} is shown the $(x_s,\sin{(\gamma)})$ plot for the
standard model with (a) present data, and (b) for the $B$ factory
cases $\sin{(2\alpha)} = 1, 0, -1$ from left to right.  $x_s$ is
determined in the SM from $x_s = 1.35 x_d (|V_{ts}|/|V_{td}|)^2$.  The
largest errors arise from the uncertainty in $|V_{td}|$, which follow
from the present $20\%$ uncertainty in $\sqrt{B_B} f_B = 200 \pm 40$
MeV from lattice calculations\cite{sharpe}.  The $B$ factory in the
SM constructs a rigid triangle from the knowledge of $\alpha$ and
$\beta$, and removes this uncertainty in $\gamma$ and $x_s$ in the
future.  A cautionary note for experiments emerges from this plot,
namely that $\sin(\gamma)$ is close to one (0.7 to 1.0) for the
1-$\sigma$ contour, and high accuracy on $\sin(\gamma)$ will be needed
to add new information to the standard model.  At 1-$\sigma$ the range
of $x_s$ in the standard model is from 14 to 33.  It is clear that the
different $\sin(2\alpha)$ cases gives distinct ranges for $x_s$.
Checking whether $x_s$ agrees with the range given by a $\sin{(2\alpha)}$
measurement will be a good test of the standard model.

\section{Constraints on the Four Down Quark Model at Present, and After the 
{\boldmath $B$} Factory Results} 

In the following, we will find and
take $\sin(2\beta) = 0.65$ as the center of the current range 
for the SM with its
projected $B$ factory errors of $\pm 0.06$ \cite{porter}, and vary
$\sin(2\alpha)$ from $-1.0$ to 1.0, using the projected $B$ factory
errors of $\pm 0.08$.

Here we also project forward to having results on $\sin{(2\alpha)}$
and $\sin{(2\beta)}$ from the $B$ factories, and show how there will
be stronger limits on the new phases of FCNC couplings than from
present data.  In the four down quark model we use
``$\sin{(2\alpha)}$'' and ``$\sin{(2\beta)}$'' to denote results of
the appropriate $B_d$ decay $CP$ violating asymmetries, but since the
mixing amplitude is a superposition, the experimental results for
these asymmetries are not directly related to angles in a triangle in
this model.  The asymmetries with FCNC contributions included are
\begin{equation}
\sin{(2\beta)} \equiv A_{B^0_d \to \Psi K^0_s} =
{\rm Im} \left[ \frac{(U^2_{std-db} + U^2_{db})}
{|U^2_{std-db} + U^2_{db}|}
\frac{(V^*_{cb} V_{cs})}{(V^*_{cb} V_{cs})^*} \right]
\end{equation}
\begin{equation}
\sin{(2\alpha)} \equiv -A_{B^0_d \to \pi^+ \pi^-} =
-{\rm Im} \left[ \frac{(U^2_{std-db} + U^2_{db})}
{|U^2_{std-db} + U^2_{db}|}
\frac{(V^*_{ub} V_{ud})}{(V^*_{ub} V_{ud})^*} \right]
\end{equation}
with $U_{std-db}$ defined in Eqn. (2.2).

In the four down quark model, what we mean by ``$\sin{(\gamma)}$'' is
the result of the experiments which would give this variable in the SM
\cite{Kayser}.  Here, the four down quark model involves
more complicated amplitudes, and $\sin{(\gamma)}$ is not simply 
$\sin{(\delta_{13})}$:
\begin{equation}
\sin(\gamma) \equiv {\rm Im}   \left[ \frac{(U^2_{std-bs} + U^2_{bs})}
{|U^2_{std-bs} + U^2_{bs}|}
\frac{(V^*_{ub} V_{cs})}{|V_{ub} V_{cs})|} \right].
\end{equation}
We note that since $\sin{(\gamma)}$ is an imaginary part of a complex 
amplitude, it can have values ranging from $-1$ to $+1$.  We now extend the
range of the previous analyses to cover the complete range.

In the four down quark model, $x_s$ is no longer the simple ratio of
two CKM matrix elements, but now involves the $Z$-mediated
annihilations and exchange amplitudes as well
\begin{equation}
x_s = 1.35 x_d \frac{|U^2_{std-bs}+U^2_{bs}|}{|U^2_{std-db}+U^2_{db}|},
\end{equation}
where
\begin{equation}
U^2_{std-bs} = (\alpha/(4 \pi \sin{\theta_W}^2))y_t f_2(y_t)
(V^*_{tb}V_{ts})^2.
\end{equation}

The asymmetry $A_{B_s}$ in $B_s$ mixing in the standard model with the
leading decay process of $b \to c \bar c s$ has no significant phase
from the decay or from the mixing which is proportional to $V_{ts}^2$.
The near vanishing of this asymmetry is a test of the standard
model\cite{nirsilnp}, and a non-zero value can result from a ``new
physics'' model.  With the FCNC, the result is
\begin{equation}
A_{B_s}  =
{\rm Im} \left[ \frac{(U^2_{std-bs} + U^2_{bs})}{|U^2_{std-bs} + U^2_{bs}|}
\frac{(V^*_{cb} V_{cs})}{(V^*_{cb} V_{cs})^*} \right]
\end{equation}
Again, since this is an imaginary part of a complex amplidute, we
extend our studies to the full range including negative values for
this.  Since it concerns the $B_s$ mixing, we plot it against $x_s$
which involves the magnitude of the amplitude used in $A_{B_s}$.

In the four-down-quark model with the unitarity quadrangle, what we
plot for the ($\rho$, $\eta$) plot is the scaled vertex of
the matrix element $V^*_{ub}$
\begin{equation}
 \rho + i \eta \equiv V^*_{ub}V_{ud}/|V_{cb}V_{cs}| .
\end{equation}
Since $\eta$ is an imaginary part, it can have negative as well as
positive values.  While the negative values were not included before
in comparing to the standard model, they are now included to show the
full range of predictions of the four-down-quark model.

We then make maximum likelihood plots which include
($\sin{(2\alpha)}$, $\sin{(2\beta)}$), ($\rho$, $\eta$), ($x_s$,
$\sin{\gamma}$), and $(x_s,A_{B_s})$.

The corresponding plots for the four down quark model are shown for
present data and for projected $B$ factory data in the following
figures.  In the figures, we show $\chi^2$ contour plots with
confidence levels (CL) at values equivalent to 1-$\sigma$ and at 90\%
CL (1.64$\sigma$) for present data, and for projected $B$ factory
results.  Again, for results with the $B$ factories, we use the
example of the most likely $\sin{(2\beta)} = 0.65$ with $B$ factory
errors of $\pm 0.06$, and errors of $\pm 0.08$ on $\sin{(2\alpha)}$.

In Fig.~\ref{rho-eta4q} we have plotted the $\chi^2$ contours for the
location of the vertex of $(\rho,\eta)$.  We note that in contrast to
the standard model, in Fig.~\ref{rho-eta4q}a the presently allowed 90\% CL
contour in the four down quark model is an annular ring representing
no constraint on $\delta = \delta_{13}$ which can result from the FCNC
with its new phases $e^{i \delta_{14}}$ or $e^{i \delta_{24}}$ in
$U_{db}$ causing the known $CP$ violation.  In Fig.~\ref{rho-eta4q}b,c
and d we show the $B$ factory cases of $\sin{(2\alpha)} = -1, 0$ and
1, respectively, with contours at 1-$\sigma$ and at 90\% CL.  The
existence of several regions, even now for negative $\eta$, requires
that extra experiments in $\sin{(\gamma)}$ or $x_s$ will also be
needed to verify or to bound out the extra down quark mixing model.
Use of the slightly more conservative bound for $|V_{ub}/V_{cb}|$
of $0.08 \pm 0.02$, which is used by some authors, still
results in multiple regions.

In order to display how the FCNC $Z^0$ exchange with the new phases
in $U_{ds}$ can account for the $CP$ violation in $\epsilon_K$, we 
plot the ratio of the FCNC contribution to the root-mean-square of
the SM and the FCNC contributions,
\begin{equation}
R^{\rm FCNC}_\epsilon = \frac{{\rm Im}(U_{ds}^2)}
{ ( (A^{\rm SM}_\epsilon)^2 + ( {\rm Im}(U_{ds}^2) )^2 )^{1/2} },
\end{equation}
so that $-1 \leq R^{\rm FCNC}_\epsilon \leq 1$.
Here $A^{\rm SM}_\epsilon = \alpha {\rm Im}(-\tilde{E}^*)/
(4 \pi \sin^2{\theta_W})$ and $E$ is from Inami and Lim\cite{inami}.
In Fig.~\ref{repsilon} $R^{\rm FCNC}_\epsilon$
is shown against the angle of $V^*_{ub}$ which is $\delta_{13}$.
In Fig.~\ref{repsilon}, for $\delta_{13}$ from $20^\circ$ to $150^\circ$, 
$R^{\rm FCNC}_\epsilon = 0$ is allowed, i.e., the SM can account for
$\epsilon_K$ in this analysis.  
At angles further outside that region, for $-180 \leq
\delta_{13} \leq 0$,
only new physics contributions can give the
imaginary part, where $R^{\rm FCNC}_\epsilon \approx 1$. 

In computing $\chi^2$ for 
a $(\sin{(2\alpha)},\sin{(2\beta)})$ contour plot for the
four down quark model we find that all pairs of
$(\sin{(2\alpha)},\sin{(2\beta)}) $ are individually allowed at
1-$\sigma$.  This
is a much broader allowed region in $\sin{(2\beta)}$ than the standard
model result from present data in Fig. 2.  The allowed 1-$\sigma$,
90\% CL  and
2-$\sigma$ contours in the ($\sin{(2\alpha)}$, $\sin{(2\beta)}$) plot
for the cases of the $B$ factory results with the four down quark
model are very similar to the SM results shown in Fig. 2.

In terms of other experiments, the $(x_s,\sin{(\gamma)}$) plot for the
four down quark model is shown in Fig.~\ref{xs-sg4q}a with the allowed
region from present data, with 1-$\sigma$ and 90\% CL contours.  This
allows all values of $\sin{(\gamma)}$ even in the extended region
from $-1 \leq \sin{(\gamma)} \leq 1$ at the 90\% CL.
At 1-$\sigma$, $x_s$ lies between 13 and 48.

In Figs.~\ref{xs-sg4q}b, c and d are shown the cases $\sin{(2\alpha)}
= -1, 0$, and 1, respectively, at 1-$\sigma$ and at 90\% CL.  They
reflect the same regions that appeared in the $(\rho,\eta)$ plots,
Figs.~\ref{rho-eta4q}b, c, and d.  The resemblance is increased if we
recall that roughly $\sin{(\gamma)} \approx \eta$, and also that $x_s
\propto x_d/|V_{td}|^2$ where $|V_{td}|$ is the distance from the
$\rho=1, \eta=0$ point.  We see that experiments on $\sin{(\gamma)}$
and $x_s$ are necessary to resolve the possible regions allowed by the
four down quark model.  For the case of $\sin{(2\alpha)}=-1$, the
allowed values of $\sin{(\gamma)}$ in Fig.~\ref{xs-sg4q}b are
different than those for the standard model in Fig.~3a.  The
$\sin{(2\alpha)}= 0$ case allows
regions of $\sin{(\gamma)}$ lower than in the SM.

The extent of the non-zero value of $A_{B_s}$ in the four down quark
model is shown in Fig.~\ref{xs-Bs4q} from present data with contours
at 1-$\sigma$, 2-$\sigma$, and 3-$\sigma$.  Plots for the $B$ factory
cases (not shown) are similar.  We note that in the new full range plot
$A_{B_s}$ is roughly symmetric about zero, with the largest absolute values
at $0.35$ at 1-$\sigma$, and $0.5 \to 0.6$ at
90\% CL.  This is much different from the $\leq 0.025$ value of $A_{B_s}$ 
in the SM.

We now report on additional plots that are not shown here.
We compared the limits on the four down quark FCNC amplitude
$|U_{db}|$ versus the standard model amplitude $|U_{std-db}|$ for
$B_d^0 - \bar{B}_d^0$ mixing, at present and after the $B$ factory
results.  At present the constraints are such that $|U_{db}|$ can go
from zero up to as large as the magnitude of $|U_{std-db}|$ at
1-$\sigma$ \cite{chosilfcnc}. 
$|U_{sb}|$ is restricted to about half of $|U_{std-bs}|$. The total phase of
$B_d^0 - \bar{B}_d^0$ mixing can range over all angles, while the
SM phase is between $-30^\circ$ and $80^\circ$ when in combination with the
FCNC amplitude.  The magnitude of $|U_{db}/(V_{cd}V_{cb})|$ in the unitarity
triangle is $\leq 0.15$ at 1-$\sigma$.

The 90\% CL limits on the three new quark mixing elements $|V_{4d}|$,
$|V_{4s}|$, and $|V_{4b}|$ are roughly equal to the mixing angles to
the fourth down quark $\theta_{14}$, $\theta_{24}$ and $\theta_{34}$,
respectively.  They are bounded by 0.05, 0.05, and 0.08, respectively.
The values allowed in combination are much more restricted, since
they are roughly bounded by hyperbolic curves,
due to constraints acting on their products in $U_{ds}$, $U_{sb}$,
and $U_{bd}$.

\section{Conclusions}

We have extended our analysis to the full range of the variables
$\eta$, $\sin{(\gamma)}$ and $A_{B_s}$, all of which are imaginary
parts, to include all of their negative values.  For the four down
quark model they all show remarkable and experimentally important new
behaviours.  From present constraints, the vertex of $V^*_{ub}$ now is
allowed in this model to be complete circular annulli about
$(\rho,\eta) = (0,0)$ at 90\% CL due to the new phases $\delta_{14}$ or
$\delta_{24}$ accounting for the presently observed $CP$ violation in
$\epsilon$.  $\sin{(\gamma)}$ is now allowed in this model over its
entire range from $-1$ to $+1$.  The range of $A_{B_s}$ is almost equally 
as large for its
negative values as it is for its positive values, and perhaps large enough 
to be observed.  Since it is
almost null in the SM, this could be a dramatic evidence of new physics.

For the $B$ factory cases there are new multifold allowed regions as
shown in the extended $(\rho,\eta)$ plots including for negative
$\eta$.  This will require additional experiments on $x_s$ and
$\sin{(\gamma)}$ to well define the four down quark model results, and
eventually to verify or bound out the relevance of the model here.  In
the $(x_s,\sin{(\gamma)})$ plot for similar cases, there are new
multiple regions for $\sin{(\gamma)}$ negative.

\acknowledgements

This research was supported in part by the U.S. Department of Energy
under Contract No. DE-FG0391ER40679.  We acknowledge the hospitality
of SLAC and CERN.  We thank Herng Tony Yao for discussions.

\begin{figure}
\caption{The $(\rho,\eta)$ plot for the standard model, showing the 1, 2, 
and 3-$\sigma$ contours, for the present data (large contours) and for
projected $B$ factory results (smaller circular contours) at
$\sin{(2\alpha)}= 1,0$, and $-1$ from left to right.}
\label{rho-etaSM} 
\end{figure}

\begin{figure}
\caption{
The ($\sin{(2\alpha)}$, $\sin{(2\beta)}$) plot for the standard model
at 1, 2, and 3-$\sigma$ with present data (nearly horizontal contours),
and with the sample results of the $B$ factories (almost circular
contours), for $\sin{(2\alpha)} = 1, 0$, and $-1$ from left to right.
\label{sa-sbSM} }
\end{figure}

\begin{figure}
\caption{
The ($x_s$, $\sin{\gamma}$) plots are shown for the standard model
with:  (a) present limits; and (b) sample results for the $B$
factories for $\sin{(2\alpha)} = 1, 0$, and $-1$ from left to right. 
\label{xs-sgSM} } 
\end{figure}

\begin{figure}
\caption{The $(\rho,\eta)$ plots for the four down quark model from:
(a) present data, and for $B$ factory cases for values of
$\sin{(2\alpha)}$ as labeled.  Contours are at 1-$\sigma$ and at 90\%
CL.
\label{rho-eta4q} }
\end{figure}

\begin{figure}
\caption{The ratio $R^{\rm FCNC}_\epsilon$ of the contribution of the 
FCNC amplitude to $\epsilon_K$ divided by the root-mean-square of the
SM and the FCNC amplitudes, as a function of the angle $\delta_{13}$.}
\label{repsilon}
\end{figure} 

\begin{figure}
\caption{The $(x_s,\sin{(\gamma)})$ plots for the four down quark
model from (a) present data, and (b, c, and d) for $B$ factory cases
for values of $\sin{(2\alpha)}$ as labeled.  Contours are the same as
in Fig.~4.
\label{xs-sg4q} }
\end{figure}

\begin{figure}
\caption{The $(x_s,A_{B_s})$ plot for the $B_s$ asymmetry $A_{B_s}$ in
the four down quark model for present data, with contours at 1, 2 and
3-$\sigma$.}
\label{xs-Bs4q} 
\end{figure}


\begin{references}
\bibitem{rbexpt} We take the 1-$\sigma$ fractional error on $R_b$ from
the deviation of the combined average from the SM result, since if there
is a real positive deviation, it would not come from FCNC but from new
physics not included here.
\end{references}
\end{document}